\documentclass[
	aps,
    prstab,
	a4paper,
	amsmath,
	amssymb,
	floatfix,
	showpacs,
	reprint,
	nofootinbib,
    ]{revtex4-2}

\usepackage[utf8]{inputenc}
\usepackage{graphicx}
\usepackage{siunitx}
\usepackage{tabularx}
\usepackage{hyperref}
\usepackage{booktabs}

\usepackage[normalem]{ulem} 

\newcommand{\figref}[1]{Fig.~\ref{#1}}
\newlength{\figwidth}
\begin{document}

\title{
     PhysBERT: A Text Embedding Model for Physics Scientific Literature
     }

\author{Thorsten Hellert}\altaffiliation{thellert@lbl.gov}
\author{João Montenegro}
\affiliation{Lawrence Berkeley National Laboratory, Berkeley 94720, California, USA}

\author{Andrea Pollastro}
\affiliation{
    Department of Electrical Engineering and Information Technology (DIETI), University of Naples Federico II, Naples, Italy\\
    Instrumentation and Measurement for Particle Accelerator Laboratory (IMPALab)
}
\date{\today}

\begin{abstract}
The specialized language and complex concepts in physics pose significant challenges for information extraction through Natural Language Processing (NLP). Central to effective NLP applications is the text embedding model, which converts text into dense vector representations for efficient information retrieval and semantic analysis. In this work, we introduce PhysBERT, the first physics-specific text embedding model. Pre-trained on a curated corpus of 1.2 million arXiv physics papers and fine-tuned with supervised data, PhysBERT outperforms leading general-purpose models on physics-specific tasks including the effectiveness in fine-tuning for specific physics subdomains.
\end{abstract}

\maketitle


\section{Introduction}  

The field of physics encompasses a vast body of knowledge, spanning numerous sub-disciplines and theoretical frameworks. The specialized language used in physics publications~\cite{Wulff_2024} and the extensive corpus of information disseminated through academic journals, textbooks, technical reports, and online repositories present significant challenges for for automated extraction of meaningful insights.

Recent advancements in Natural Language Processing (NLP) are fundamentally transforming our ability to analyze and process textual data~\cite{min2021recentadvancesnaturallanguage}. At the forefront of this transformation are text embedding models~\cite{mikolov2013efficientestimationwordrepresentations,peters_2008_deep}, which convert textual data into dense vector representations, capturing its semantic meaning and enabling computational analysis such as efficient information retrieval~\cite{manning2008introduction}, text classification~\cite{howard_ruder_2018_universal}, and semantic similarity measurement~\cite{reimers-2019-sentence-bert}. In academic research, domain-specific embeddings can significantly enhance the accuracy of literature reviews by clustering related papers~\cite{Singh2022SciRepEvalAM}, identifying emerging trends~\cite{Sulc2024}, and improving the precision of reviewer matching tools for scientific journals~\cite{zhang2024chainoffactors}.

In the last few years,  Transformers~\cite{wolf2020huggingfacestransformersstateoftheartnatural} have become the foundation of these models~\cite{devlin2019bertpretrainingdeepbidirectional,reimers-2019-sentence-bert} owing to their self-attention mechanism, which has significantly enhanced context awareness in NLP tasks. Building on this foundation, Large Language Models (LLMs), such as GPT~\cite{openai2024chatgpt} and LLaMA~\cite{llama3}, have further advanced the field of NLP by providing powerful tools for understanding and generating human language, thereby facilitating the extraction of meaningful insights from complex textual data. However, these models often suffer from hallucinations~\cite{10_1145_3571730}, producing plausible-sounding but incorrect or nonsensical information. To address this issue, Retrieval-Augmented Generation (RAG)~\cite{lewis2021retrievalaugmentedgenerationknowledgeintensivenlp} has surged in popularity~\cite{gao2024retrievalaugmentedgenerationlargelanguage}, as it combines the generative capabilities of LLMs with the precision of retrieval systems. Central to the effectiveness of RAG pipelines is the text embedding model, which plays a crucial role in matching queries with source documents, thereby ensuring the precision and relevance of the retrieved information, with specialized embedding models proving superior to general ones~\cite{beltagy2019scibert}.

General-purpose text embedding models~\cite{huggingface_mteb_leaderboard}, typically trained on a diverse range of internet texts~\cite{muennighoff2023mtebmassivetextembedding}, lack the specialized knowledge required to accurately represent the language of specific disciplines. Specialized embedding models have demonstrated significant improvements across various fields in natural science, including chemistry~\cite{shermukhamedov2023structurepropertychemicalelement}, material science~\cite{Gupta2022MatSciBERT}, and the biomedical domain~\cite{1093_bioBERT}. However, the domain of physics notably lacks embedding models specifically tailored to its unique semantic characteristics. Consequently, general-purpose embedding models are currently utilized in physics NLP applications due to the absence of specialized alternatives~\cite{hexemer2024beamline,Sulc_2024ssg,Rehm2024,Murnane2024,Steinbach2024}.

\begin{figure*}[htb]
    \centering
    \includegraphics[width=1\linewidth]{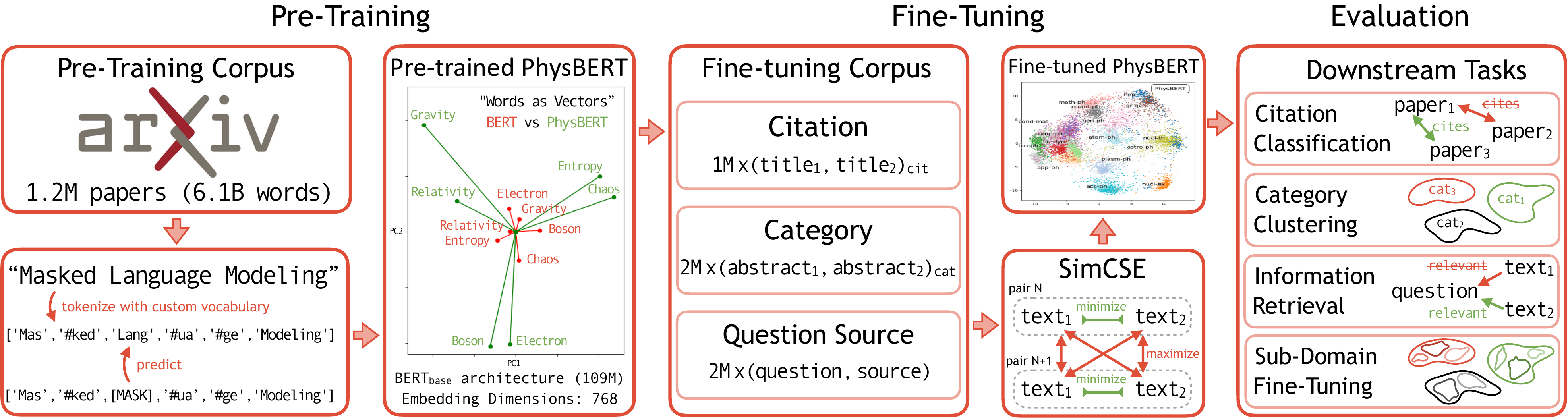}
    \caption{Schematic overview of the steps involved in developing PhysBERT. The process begins with pre-training on a large corpus from arXiv, followed by supervised fine-tuning using SimCSE. Finally, the model is evaluated on downstream tasks such as citation classification, category clustering, information retrieval, and sub-domain fine-tuning.}
    \label{fig:overview}
\end{figure*}

In this context we introduce PhysBERT, a sentence embedding model specifically designed for the field of physics. Leveraging the BERT~\cite{devlin2019bertpretrainingdeepbidirectional} architecture, PhysBERT is trained on a curated corpus of physics literature based on 1.2 million physics papers available on arXiv~\cite{arxiv}, encompassing a wide range of sub-disciplines within the field. To validate the effectiveness of PhysBERT, we create specific datasets and downstream evaluation tasks such as information retrieval, classification, and semantic similarity, all tailored to the physics domain. The combination of comprehensive pre-training and targeted, supervised fine-tuning equips PhysBERT with a deep understanding of physics language, enabling it to significantly outperform general-purpose models on these physics-related NLP tasks. Additionally, we demonstrate that PhysBERT serves as an excellent starting point for fine-tuning in specific physics subdomains, highlighting its adaptability and potential for further specialization.

A schematic overview of the workflow described in this paper is provided in \figref{fig:overview}. Section~\ref{sec:training} details our pre-training and fine-tuning methodology, while Section~\ref{sec:downstream} covers the downstream tasks used for model evaluation. The datasets developed for training and evaluation are introduced in Section~\ref{sec:data}. Finally, we present the experimental setup and results in Section~\ref{sec:results}. In addition to our model weights, we are releasing the training and evaluation datasets alongside this manuscript~\cite{HF_physbert}.


\section{Training}\label{sec:training}

A preliminary step in training our embedding model involves the building of a tokenizer. A tokenizer plays a fundamental role in text processing, as it converts text into integers (tokens) that the model can effectively handle. Given our extensive dataset of 40GB of text (see Section~\ref{sec:data}), which provides the capacity to train a new model from scratch, we build a custom tokenizer, following the BERT~\cite{devlin2019bertpretrainingdeepbidirectional} approach with the standard vocabulary size of 30,523. The full training process is carried out for both a cased and an uncased version of the model. In the cased version, the tokenizer preserves the capitalization of words, which can be important for capturing nuances in meaning or acronyms, while in the uncased version, all text is converted to lowercase, simplifying the model’s learning task. 

We begin our training process with unsupervised learning, utilizing the BERT\textsubscript{base} architecture~\cite{devlin2019bertpretrainingdeepbidirectional}, initialized with random weights. It is important to note that BERT exists in two variants: BERT\textsubscript{base} (109M model parameters) and BERT\textsubscript{large} (335M parameters), with the latter offering increased capacity at the cost of significantly higher computational demands. Due to these substantial resource requirements at both training and inference, our studies are confined to the BERT\textsubscript{base} variant, which results in an embedding space with a dimensionality of 768. 

Pre-training adheres to the RoBERTa methodology~\cite{liu2019robertarobustlyoptimizedbert}, focusing exclusively on Masked Language Modeling (MLM)~\cite{devlin2019bertpretrainingdeepbidirectional}. In MLM, the model is trained to predict missing words within a text sequence. Typically \SI{15}{\%} of input tokens are randomly replaced with a [MASK] token. The model’s objective is to accurately predict the original words based on the surrounding context, allowing it to learn deep bidirectional text representations by considering both preceding and succeeding words.

The pre-training phase involves two steps: first, MLM training with a sequence length of 128 tokens, followed by an extension to 512 tokens to capture longer dependencies. When starting from random weights, this incremental approach is more robust and enables faster convergence then starting directly with the full context length of 512. Pre-training is conducted over 10 epochs with a batch size of 8, an MLM probability of \SI{15}{\%}, and a learning rate of 1E-4, using Adam~\cite{lecun2015deep} as the optimizer.

As illustrated in \figref{fig:overview}, the pre-trained model demonstrates a significantly improved ability to distinguish and group physics-related keywords compared to general-purpose text embedding models. However, a notable limitation of the BERT network structure is that it does not produce independent sentence\footnote{It is important to note that in the context of language models, ‘sentences’ refer not only to individual sentences but to larger text segments as well.} embeddings, making it difficult to obtain semantically meaningful sentence representations directly from BERT~\cite{reimers-2019-sentence-bert}. In response to these limitations, we fine-tune~\cite{Ding2023} our model using Simple Contrastive Learning of Sentence Embeddings (SimCSE)~\cite{gao2022simcsesimplecontrastivelearning} in the final stage of our training process within the Sentence Transformer~\cite{reimers-2019-sentence-bert} framework. SimCSE is an efficient contrastive learning method that enhances the model’s ability to produce semantically meaningful sentence representations by minimizing the distance between positive pairs (similar sentences) and maximizing the distance between negative pairs (dissimilar sentences).

Given adequately structured input data, this results in more precise and meaningful sentence representations, which are essential for downstream tasks that require a high level of semantic understanding. For this supervised training setup our data consist of semantically similar sentence pairs as described in Section~\ref{sec:sub_data}, with all other sentences in the batch treated as negatives.


\section{Downstream tasks}\label{sec:downstream}

To evaluate the effectiveness of our fine-tuned embedding model, we conduct comprehensive benchmark tests across various downstream tasks. These diverse evaluations measure the models’ performance in practical applications. Due to the lack of publicly available benchmarks for scientific physics publications, we developed a custom set of assessments, closely adhering to the methodologies in recognized text embedding benchmarks~\cite{muennighoff2023mtebmassivetextembedding, thakur2021beirheterogenousbenchmarkzeroshot}.

\subsection{Category Clustering} 

Clustering is a robust method for evaluating how well a model uncovers the inherent semantic structure of a dataset without predefined labels. By organizing sentences into coherent categories, the benchmark demonstrates the model’s ability to generalize across various topics and contexts, crucial for effective topic modeling~\cite{grootendorst2022bertopicneuraltopicmodeling, sulc2023textualanalysisicalepcsipac,Sulc2024,CHAGNON2024100044}. Clustering also directly assesses the quality of the model’s embedding space, as successful clustering relies on a well-structured, meaningful text representation (see \figref{fig:clustering}).

The inputs for this benchmark are sentences paired with their ground truth labels, indicating the physics category each sentence belongs to. Sentences within the same category should cluster together as they address the similar topics. First, the sentences are embedded into vector representations. Then, the KMeans~\cite{pedregosa2018scikitlearnmachinelearningpython} algorithm is used to group the embeddings into clusters, with the number of clusters matching the number of unique labels in the dataset. Clustering performance is assessed using the V-measure score~\cite{rosenberg_hirschberg_v}, evaluating both homogeneity and completeness. To ensure robust and reliable evaluation, we utilized a stratified 10-fold cross-validation~\cite{rosenberg-hirschberg-2007-v}. Each fold involves splitting the data into training and test sets, standardizing the training set, and fitting a KMeans model. The final performance metric is the mean V-measure score across all the test sets.

\subsection{Information Retrieval} 

Information retrieval~\cite{Hambarde_2023} is a pivotal downstream task in the context of RAG applications, as it underpins the system’s ability to fetch relevant information from vast corpora of documents. Effective information retrieval enhances the accuracy and relevance of the generated responses, thereby improving the overall performance of RAG systems.

In order to robustly evaluate the model performance, we follow common information retrieval benchmarking practices~\cite{thakur2021beirheterogenousbenchmarkzeroshot,muennighoff2023mtebmassivetextembedding}. Each dataset includes a corpus of documents, a set of queries, and a mapping that links each query to its relevant documents. The goal is to accurately retrieve the relevant documents for a given query. Following standard RAG procedures, the embedding model transforms all queries and documents into embeddings, and cosine similarity scores are calculated between each query and all documents. Documents are then ranked based on these scores. Retrieval effectiveness is measured using the normalized Discounted Cumulative Gain at rank 10 (nDCG@10)~\cite{wang2013theoreticalanalysisndcgtype}.

\subsection{Citation Classification}

Citation classification is a valuable task for a scientific embedding model because it demonstrates the model’s ability to understand and represent the nuanced relationships between scientific papers. A proficient model can aid in recommending relevant literature, identifying emerging research trends, and discovering implicit connections between works that may not directly cite each other. These capabilities enhance the depth and breadth of literature reviews~\cite{VANDINTER2021106589}, improve the precision of research impact analysis~\cite{Singh2022SciRepEvalAM}, and support the organization of scientific knowledge~\cite{semanticscholar}.

To evaluate the embedding models on this task, we use a binary classification benchmark~\cite{reimers-2019-sentence-bert} with a dataset of paper title pairs—some that cite each other and some that don’t. The process begins by generating embeddings for each title. The goal is to measure the similarity between these embeddings to classify the pairs as citing or non-citing. We use a balanced dataset with equal numbers of positive (citing) and negative (non-citing) pairs. Similarity is measured using cosine similarity, and pairs are classified by identifying the optimal threshold separating positive and negative labels. The model’s accuracy, referred to as cosine accuracy~\cite{singhal2001modern}, is calculated based on the percentage of correct classifications.

\subsection{Fine-tuning on physics subdomains}

Developing a sentence embedding model using extensive datasets, such as all arXiv physics publications, enables us to capture the broad and diverse nature of the field, providing a solid foundation for various applications. However, to achieve optimal performance in specific subdomains, fine-tuning the model on targeted subsets of data will likely become essential for future applications.

To demonstrate the effectiveness of PhysBERT as a foundation for domain-specific fine-tuning, we leverage the extensive nature of three large categories within arXiv—Condensed Matter, Astrophysics, and High Energy Physics—each of which comprises multiple subcategories. For instance, Astrophysics includes explicit subcategories such as ‘Cosmology and Nongalactic Astrophysics’ and ‘Earth and Planetary Astrophysics’ (see Ref.~\cite{arxivCategoryTaxonomy} for all categories).

For the evaluation of this fine-tuning task we use a simplified setup akin to the supervised fine-tuning setup described above, with category clustering as the only evaluation metric.


\section{Datasets}\label{sec:data}

Carefully curated training data is essential for developing accurate and robust language models. To this end, we have created various datasets tailored to our study’s needs. We differentiate between datasets used for unsupervised pre-training and those employed for supervised fine-tuning.

\subsection{Unsupervised pre-training}
For unsupervised pre-training, we utilize an extensive corpus of text compiled from scientific publications. We download all available papers from arXiv~\cite{arxiv}, including both PDFs and the available metadata using the provided bulk data access~\cite{bulk_data_s3}. We restrict the postprocessing to papers categorized by their authors under one of the 61 physics categories~\cite{arxivCategoryTaxonomy}, which totalled to 1,25 million papers.

All PDFs are processed using the Optical Character Recognition (OCR) tool Nougat~\cite{blecher2023nougatneuralopticalunderstanding}, which converts the paper text into markdown format. For training, we utilize a postprocessed version containing only the full text of the sections, excluding captions, references, and any content preceding the abstract or following the references section. Following the methodology outlined in Ref.~\cite{liu2019robertarobustlyoptimizedbert}, we concatenated all clean text from the documents, resulting in a corpus comprising 41 GB of text or about 6.1B words.

\subsection{Supervised fine-tuning}\label{sec:sub_data}
Supervised learning in the context of sentence embeddings involves distinguishing between similar and dissimilar pairs of text (see Sec.~\ref{sec:training}). We identified several datasets that can be utilized for this purpose.

\paragraph{\textbf{Abstract pairs from categories:}}
ArXiv publications are categorized based on the primary category assigned by the authors upon submission. Recognizing that papers within the same category are likely to be contextually more similar than those from different categories, we leverage this structure to create our dataset of paired abstracts. To ensure robustness, we exclude categories with fewer than 5,000 papers and combine all subcategories under Astrophysics, Condensed Matter, and High Energy Physics—categories so extensive that they have subcategories—into their respective main categories. This approach leaves us with 21 categories, from which we draw 2 million abstract pairs, equally distributed across the categories to ensure a balanced dataset. 

\paragraph{\textbf{Citation pairs:}}
Citations are a valuable piece of information in large publication databases, providing insights into the contextual relationships between papers. It can be assumed that papers citing each other are contextually closer than those that do not. To leverage this information, we build a comprehensive citation tree using the Semantic Scholar~\cite{semanticscholar} database API to query the references of papers in our arXiv database. By doing so, we can identify and pair the titles of papers that cite each other. We include 1M citation pairs in the training set.

\paragraph{\textbf{Synthetic Query-Source Data:}}\label{sec:synthentic_data}
One particularly valuable form of annotated training data is query-source pairs, where a query is linked to the text source containing the answer. These pairs are essential for enhancing a model’s ability to handle inquiries which is critical for any RAG application~\cite{thakur2021beirheterogenousbenchmarkzeroshot}. However, in specialized fields such as physics, acquiring vast amounts of annotated query-source data poses a significant challenge due to the scarcity of such datasets.

To address this limitation, we use data augmentation, which artificially creates data to mimic real-world characteristics and patterns rather than directly collecting it~\cite{liu2024bestpracticeslessonslearned}. In our study, we generate question-and-answer pairs from text chunks extracted from research papers, following a setup similar to standard RAG workflows~\cite{gao2024retrievalaugmentedgenerationlargelanguage}. Specifically, we select a random 1000-character text chunk from a research paper and prompt a locally running LLM to generate three question-and-answer pairs that can be exclusively answered by the provided text, with the abstract given as general context.

In order to generate 2M questions-source pairs, we use LLaMA3-70B~\cite{llama3}, a notably large model requiring substantial computational resources for processing which constraints the amount of generated text significantly. However, our experiments with smaller models produced lower-quality text and a high rate of nonsensical questions, especially when the source text included substantial mathematical content.

\subsection{Model evaluation data}\label{sec:data_model_eval}

The model evaluation uses datasets tailored to each downstream task.

For clustering in general physics, the input data includes 1,000 labeled paper abstracts from each of 21 major physics categories on arXiv. For citation classification, we utilize 50k pairs of paper titles that cite each other and 50k randomly drawn non-citing pairs. For information retrieval, we use 50k query-source pairs as described above. In all cases, the evaluation data is kept separate from the training sets.

For subsequent subdomain fine-tuning, we focus on the three largest arXiv physics categories with explicit subcategories: Condensed Matter (10 subcategories), Astrophysics (7 subcategories), and High Energy Physics (4 subcategories). We use this substructure to build category-based datasets for both supervised fine-tuning and clustering evaluation, following the corresponding approaches outlined above. The fine-tuning training data consists of 10k abstract pairs per subcategory, while the clustering evaluation uses 1k labeled abstracts per subcategory, not included in the training set.


\section{results}\label{sec:results}

\begin{figure*}
    \centering
    \includegraphics[width=0.9\linewidth]{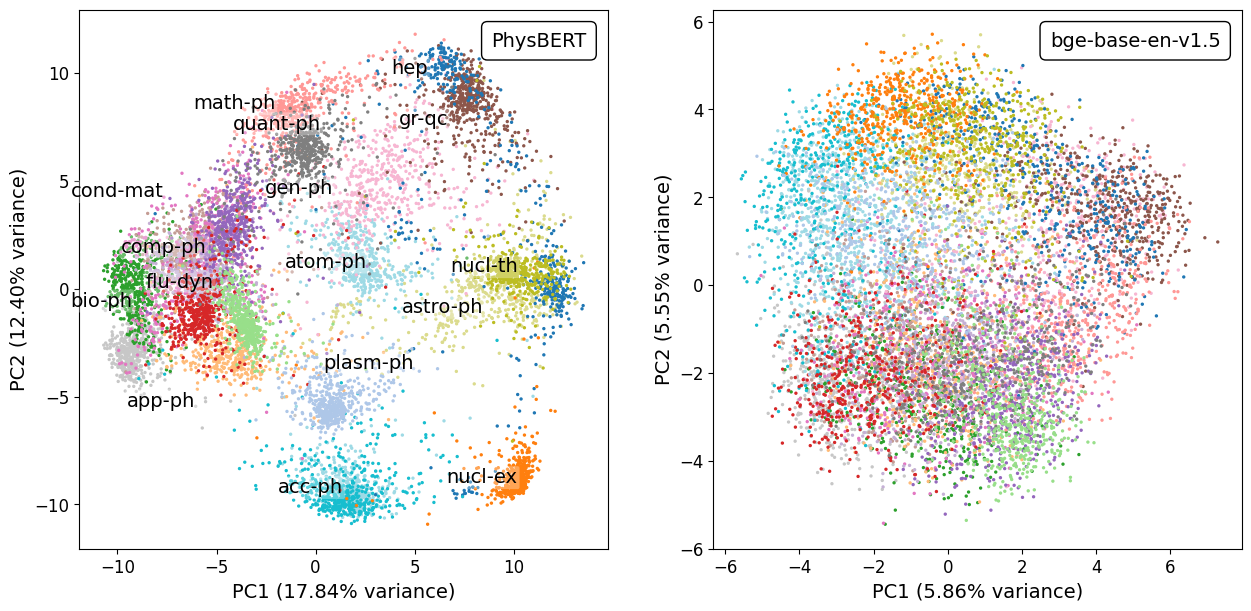}
    \caption{Comparison of embedding space visualizations for PhysBERT (left) and bge-base-v1.5~\cite{bge_embedding} (right, see also Table~\ref{tab:results}), using PCA on text embeddings from 500 random abstracts per physics category. It is worth noting that no explicit clustering algorithm was applied; the observed patterns reflect the model’s internal organization of the data.}
    \label{fig:clustering}
\end{figure*}

We pre-train our model on 32 nodes each containing 4 NVIDIA A100 GPUs at the National Energy Research Scientific Computing Center (NERSC) \cite{NERSC}, utilizing PyTorch in Distributed Data Parallel mode \cite{li2020pytorchdistributedexperiencesaccelerating}. The training process is conducted over a total of four epochs with a sequence length of 128 tokens and six epochs with a sequence length of 512 tokens. For supervised fine-tuning, given that only two epochs are required to achieve sufficient convergence, we conduct the training on eight A100 nodes. We utilize cached gradients~\cite{gao2021scalingdeepcontrastivelearning} to effectively manage memory usage for large batch sizes~\cite{wang2024textembeddings}, in our case 256 (per gpu).

During fine-tuning, the models are evaluated on all downstream tasks described in Section~\ref{sec:downstream} three times per epoch. 
Following an hyperparameter exploration, the learning rate was set to 2E-4, the batch size to 256, the SimCSE temperature to 0.05, and the weight decay to 0.01, using Adam as optimizer.
We chose the model that yields the best overall performance across the three evaluation metrics and compare it against several models that are of particular interest to the physics community~\cite{hexemer2024beamline,Sulc_2024ssg,Rehm2024,Murnane2024,Steinbach2024}, as well as leading models from the MTEB leaderboard~\cite{huggingface_mteb_leaderboard} that are derived from the BERT\textsubscript{base} model. Notably, we include four models, derived from the BERT\textsubscript{large} architecture, which are the top performers on the MTEB leaderboard in their corresponding model parameter size of 335M.

The results in Table~\ref{tab:results} show that PhysBERT outperforms existing models, achieving the highest V-measure scores for clustering, maximum cosine accuracy for citation classification, and normalized Discounted Cumulative Gain at rank 10 (nDCG@10) for information retrieval. In particular, PhysBERT also outperforms the leading models with significantly larger parameter sizes, underscoring its impressive efficiency and superiority in handling complex physics-related NLP tasks despite its comparatively smaller size. A graphical representation of the embedding space can be found in \figref{fig:clustering}, where we project the 768 dimensional embeddings of 500 random abstracts of each physics category into two dimensions using Principal Component Analysis (PCA). This visualization reveals the model’s internal state, demonstrating its ability to distinguish between different fields within physics by clustering related categories together while separating those with distinct thematic content.

\begin{table}[h]
\centering
\caption{Downstream task results for various (uncased) text embedding models. Reported values are average V-measure score for category clustering, cosine accuracy score for citation classification, and normalized Discounted Cumulative Gain at rank 10 (nDCG@10) for the information retrieval evaluation. The top section lists text embedding models based on BERT\textsubscript{base} with 109M parameters, including our model pre-trained on MLM only and the bottom section features BERT\textsubscript{large}-based sentence embedding models with 355M parameters.}
\begin{tabular}{|l|c|c|c|}
\hline
 & \textbf{Cit.Class.} & \textbf{Cat.Clust.} & \textbf{Inf.Retr.} \\ 
\hline
BERT\cite{devlin2019bertpretrainingdeepbidirectional} & 72.4  & 36.4  & 2.1   \\
bge-base-v1.5\cite{bge_embedding}                     & 89.5  & 58.1  & 28.2  \\
E5-base\cite{wang2024textembeddings}                  & 83.4  & 54.8  & 26.1  \\
MiniLM-L6-v2\cite{allminilmL6v2}                      & 84.1  & 54.6  & 19.9  \\
mpnet-base\cite{allmpnetbasev2}                       & 85.3  & 57.4  & 23.1  \\
PACuna\cite{sulc2023pacunaautomated}                  & 74.6  & 28.5  & 3.9   \\
RoBERTa\cite{liu2019robertarobustlyoptimizedbert}     & 64.8  & 33.1  & 0.2   \\
SciBERT\cite{beltagy2019scibert}                      & 75.5  & 44.8  & 1.7   \\
SPECTER2\cite{Singh2022SciRepEvalAM}                  & 83.4  & 52.0  & 3.9   \\
PhysBERT$_\text{MLM}$ (ours)                    & 56.1  & 49.1  & 1.0   \\
PhysBERT (ours)                                       & \textbf{94.7}  & \textbf{88.9} & \textbf{35.0} \\
\hline
\textbf{Large Models} & & & \\
E5-large\cite{wang2024textembeddings}                 & 84.9  & 56.8  & 30.6  \\
UAE-Large-V1\cite{li2024angleoptimizedtextembeddings} & 89.7  & 58.3  & 29.7  \\
mxbai-large-v1\cite{emb2024mxbai}                     & 89.7  & 58.2  & 29.0  \\
bge-large-v1.5\cite{bge_embedding}                    & 89.6  & 58.3  & 29.6  \\
\hline
\end{tabular}
\label{tab:results}
\end{table}

Finally, we aim to test the ability of different models to be fine-tuned on three physics subdomains. Each model is trained for 1 epoch using a linear learning rate decay schedule. To ensure fair comparisons across different models, we perform a grid search to optimize the learning rate and the batch size within the search spaces \{1E-4, 2E-4\} and \{16, 32\}, respectively, for each training run. Throughout the training process, we evaluate the performance on category clustering three times and identify the model checkpoint with the highest average V-measure score and report the results in Table~\ref{tab:fine-tuning}. Our fine-tuned PhysBERT outperformed other fine-tuned reference models, achieving the highest average V-measure across all categories. While arguably a simplified experimental setup with only one clustering task, it underscore PhysBERT’s potential as a robust foundation for future domain-specific applications. Notably, PhysBERT\textsubscript{MLM}, the version of our model that is only pre-trained on MLM, outperforms even the larger reference models as well. This result illustrates that unsupervised pre-training on a large corpus of physics publications including a domain specific vocabulary provides a strong foundation for subsequent fine-tuning on specialized tasks.

\begin{table}[h]
\centering
\caption{Average V-measure scores for category clustering evaluation of models fine-tuned in the physics subdomains Condensed Matter (10 subcategories), Astrophysics (7 subcategories) and High Energy Physics (4 subcategories) and their average performance.}
\begin{tabular}{|l|c|c|c|c|}
\hline
{} &  \textbf{Cond.Mat.} &  \textbf{Astroph.} &   \textbf{HEP} &  \textbf{Avg.} \\
\hline
BERT\cite{devlin2019bertpretrainingdeepbidirectional} & 58.4  & 65.7  & 81.9  & 68.6  \\
bge-base-v1.5\cite{bge_embedding}                     & 60.0  & 67.5  & 84.9  & 70.8  \\
E5-base\cite{wang2024textembeddings}                  & 58.7  & 67.3  & 82.8  & 69.6  \\
MiniLM-L6-v2\cite{allminilmL6v2}                      & 54.9  & 63.6  & 80.2  & 66.2  \\
mpnet-base\cite{allmpnetbasev2}                       & 57.1  & 65.8  & 83.1  & 68.7  \\
PACuna\cite{sulc2023pacunaautomated}                  & 58.2  & 65.8  & 82.4  & 68.8  \\
RoBERTa\cite{liu2019robertarobustlyoptimizedbert}     & 55.5  & 64.9  & 80.4  & 66.9  \\
SciBERT\cite{beltagy2019scibert}                      & 59.7  & 66.4  & 85.0  & 70.4  \\
SPECTER2\cite{Singh2022SciRepEvalAM}                  & 60.0  & 67.2  & 85.0  & 70.7  \\
PhysBERT$_\text{MLM}$ (ours)                    & 60.4  & 67.9  & 86.2  & 71.5  \\
PhysBERT (ours)                         & \textbf{67.6} & \textbf{70.4} & \textbf{88.3} & \textbf{75.4} \\
\hline
\textbf{Large Models} & & & & \\
E5-large\cite{wang2024textembeddings}                 & 59.9  & 68.3  & 84.1  & 70.8  \\
UAE-Large-V1\cite{li2024angleoptimizedtextembeddings} & 60.3  & 68.0  & 85.0  & 71.1  \\
mxbai-large-v1\cite{emb2024mxbai}                     & 59.9  & 68.1  & 84.5  & 70.8  \\
bge-large-v1.5\cite{bge_embedding}                    & 60.1  & 67.9  & 84.1  & 70.7  \\
\hline
\end{tabular}
\label{tab:fine-tuning}
\end{table}

\section{Conclusion}

In this work, we have introduced PhysBERT, the first sentence embedding model specifically trained on scientific publications within the field of physics. Our approach began with the development of a custom tokenizer optimized for the physics domain, followed by pre-training on an extensive dataset of 1.2 million arXiv papers. This initial training was complemented by SimCSE fine-tuning on curated datasets, including data synthetically generated by an LLM, to enhance the model’s contextual understanding. We established four distinct evaluation metrics for downstream tasks: physics category clustering, information retrieval, citation classification, and the model’s ability to be fine-tuned further on specific physics subdomains. Our evaluation shows that PhysBERT significantly outperforms existing models across all metrics.

\section{Acknowledgments}
The authors would like to express their gratitude to the NERSC team for their exceptional user support. Their dedication, patience and prompt responsiveness were instrumental in facilitating our computational endeavors, ensuring smooth resolution of any issues encountered.

Work supported by the Director of the Ofﬁce of Science of the US Department of Energy under Contract no. DEAC02-05CH11231.
This research used resources of the National Energy Research Scientific Computing Center (NERSC), a Department of Energy Office of Science User Facility using NERSC award ERCAP0027412.

\end{document}